# MARKET DIGITALIZATION AND RESILIENCE IN CRISIS TIMES

Guillermo J. Larios-Hernandez, Universidad Anahuac Mexico, guillermo.lariosh@anahuac.mx

**Abstract:** Based on a conceptual framework that integrates three dimensions of digital transformation (DT), namely, the nature of the product, client interaction, and the level of coordination with industry players, this paper aims to explain the level of influence that contextual crisis factors may have played in organizational digitalization choices in search for resilience as part of adaptation strategies. In particular, this investigation would analyze digitalization choices as survival strategies for COVID-19 crisis in the case of Mexican enterprises. The selected country is of particular interest as research target in the Global South, in that public policy has offered little support to keep business organizations up and running, leaving entrepreneurs with no other option but to implement bottom-up resilience strategies, including digitalization. Qualitative Comparative Analysis (QCA) has been proposed to identify combinations of conditions to explain the role played by COVID crisis-related contextual factors that may have led to particular forms of digitalization. Semi-structured interviews with industry associations are also proposed to gain knowledge about group responses to the crisis.

## 1. INTRODUCTION

We are living in a time of significant turmoil, especially considering that our world societies have been heavily hurt by the COVID-19 pandemic, leaving governments and the society at large with little action choices to what appears to be a significant systemic failure. Under these circumstances, many governmental regimes appear to prioritize short-term commitments but failed to assess the long-term consequences of their policy decision-making, suggesting the existence of a dichotomy between urgent demands of "situation" and sustainable recommended paths. As participants of a complex socioeconomic system, organizations need to confront contradictory institutional signals, triggering distributed social involvement when top-down institutional and market-based mechanisms seem ineffective to members in a given society.

On the other hand, characterized by complexity, uncertainty and appraising dimensions, scholars recognize that innovative solutions to grand challenges require collective action, whose systemic linkages are still to be understood (Ferraro et al, 2015). Given the upsurge of the digital economy as the dominant techno-economic paradigm of our time (Perez, 2003), innovation can be hardly appreciated without considering the role of information and communication technologies (ICT). The fact that COVID crisis has forced wide digitalization in practically all types of organizations (Klein & Todesco, 2021) is a clear exemplification of ICT having such an instrumental role, driving organizational change to what appears to be another level of business venture evolution, whose transformational impact would depend on the degree of technological integration that a given organization decides to accomplish in its development strategy. This form of digital upheaval is the so-called digital transformation (DT).

The promise of digital transformation has encouraged scholars and think tanks to derive explanatory models and theoretical frameworks that conceptualize DT's core attributes and level of impact on several forms of organization (Andal-Ancion, Cartwright & Yip, 2003; Andriole, 2017; Gray, El

Sawy, Asper, & Thordarson, 2013; Matt et al., 2015; Schwertner, 2017; Resca, Za, & Spagnoletti, 2013; Sanchez, 2017). However, while DT is about creating new value and more revenue that originate in digitalization (Singh and Hess, 2017), approaches to explain the concept diverge, posing several problems to determine the extent of DT and its ultimate business implications. Yet, several scholars coincide in identifying DT with digital product innovation and client experience (Andal et al., 2003; Sanchez, 2017; Schwetner, 2017; Sebastian, Ross, Beath, Mocker, Moloney, & Fonstad, 2017), with deep implications on the organization's product portfolio and business model (Matt, Hess, & Benlian, 2015).

Hence, scholars continue to work on new research and case studies, each of which derives in academic contributions that offer aggregates of distinctive qualities to characterize DT from a particular perspective, defining different characters of DT that can be applied to change business organizations. But change can also drive organizational resilience and, though the principles of digital transformation usually encourage the business organization to drive both, cost down (optimization) and product differentiation up (innovation) (Sanchez, 2017), digitalization allows also for the development of new strategic alternatives to act upon external chocks and crisis, pointing to the necessity of determining the product portfolio suitability in the face of grand challenges, particularly in the case of COVID-19. Based on the three categories proposed by Andal-Ancion et al. (2003) to deploy a DT strategy, this paper aims to explain the level of influence that contextual factors may have played in organizations to digitalize, particularly, in terms of changing the nature of the product, client interaction, and the level of coordination with industry players, who seek to gain resilience as part of their adaptation strategies. The identification of such DT influencers has practical implications in that they may help scholars and practitioners in their pursuit to determine contextual factors in COVID crisis time that have led to certain DT resilience decisions.

Based on data gathered from the Mexican National Institute of Statistics (INEGI) and interviews with industry associations, this proposal aims to explain the level of influence that contextual factors may exert on the three categories that constitute DT. Mexico is particularly interesting as research target in the Global South, in that its current administration has provided little support to keep business organizations up and running. As of today, more than one million businesses have gone bankruptcy in the country and, considering that small and medium-sized enterprises (SMEs) tend to be the most vulnerable organizations in crisis situations – as a result of their limited financial resources and knowledge – (Klein & Todesco, 2021), market demand losses leave existing entrepreneurs with the sole alternative to adopt bottom-up strategies, including digitalization.

## 2. ANALYTICAL PERSPECTIVES OF DIGITAL TRANSFORMATION

Based on the previous discussion, we ought to acknowledge that the transformational effects on the business organization go well beyond the automation and optimization of business processes. To begin, the exploitation of the potential propounded by digital transformation is directly related to the type of product portfolio, which indicates those aspects on which a business organization must work to transform their business offering. Product portfolio digitalization determines to a greater extent the value proposition of the organization, which in turn shapes the strategic alternatives to implement DT. One of the key objectives of digital transformation is the embeddedness of digital technologies in products, services and business models in order to reach higher value-added in the portfolio. Additionally, digital solutions are suitable as relational interfaces that complement or substitute existing channels. These are common reasons to assimilate digitalization in business operations and new product development strategies. From a product portfolio perspective, however, Andal-Ancion et al. (2003) suggests that effective digital transformation depends on three main categories (object of impact): the nature of the product, the interaction with customers, and the level of collaboration with industry players, as shown in Figure 1.

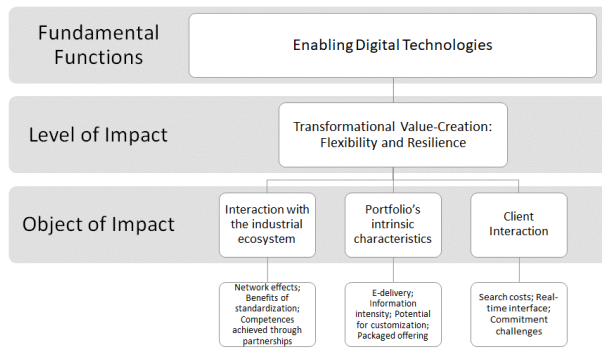

**Figure 1. Product portfolio categories that define digital transformation. Based on Andal-Ancion et al. (2003)**

Regarding the first category, that is, the inherent characteristics of a product or service, the business organization would need to assess the extent to which its product portfolio can be delivered electronically, the information intensity implicit in the products, potential for customization, and the implications of packaging products and services. The second category involves the interactions between the business organization and its customers, considering improvements to the client experience in attributes such as product portfolio's search costs, real-time interface, and a decrease in the complexity of contracting or purchasing. In the end, a new customer experience is achieved particularly through multichannel interactions (Singh and Hess, 2017). Lastly, the third category considers interactions between the organization and its partners, which aim to generate greater network effects that lead to a larger client adoption rate, take advantage of the benefits of value chains through standardization or complement the lack of in-house skills through business alliances. Digital tools would allow for a more effective management of the organization's value networks. Coincidently, Rogers (2016) names value networks a key part of an organization's business model.

As a result of technology diffusion among the population, ICT efforts in organizations move away from the enterprise core (the center) to focus on digitizing customer interaction (the edge), taking advantage of the client's digital connectivity to increase information exchange and transactions (Gray et al., 2013). Gray et al. (2013) also indicate that this customer interaction (the edge) has the potential to transform the enterprise core processes (the center) as clients can interact directly with the core because of a demand-pull trend. Thus, the analysis of digital transformation from a center-edge approach encourages the business organization to review existing ICT practices related to client interaction. Is the organization taking advantage of digital technologies to advance pull-mode interactions that start from what the client determines as valuable? Or is the ICT focus stagnant at the core, trying to push clients towards certain business ends?

However, the transformational value that can be obtained from DT (level of impact) depends not only on the functions of a generic technology, e.g. enabling digital technologies, but also on common organizational principles that imply a sociotechnical perspective, in that technology needs to be integrated into the various processes of the organization (Yunis, Tarhini, & Kassar, 2018), recognizing that technology cannot be simply technical or social, but the outcome of socio-material routines created through a process of interlacing the material nature of the innovation and the social practice. This perspective demands the business organization to carry out an in-depth analysis of what represents value for its clients, not from the ICT viewpoint but especially, from those responsible for the business results. This approach would allow business organizations to determine the structures that make sense to create a distinction in value proposition that dramatically supersedes the value offered by their competitors (Rogers, 2016), leading the business to a true digital transformation of the business model.

In coincidence with Andal-Ancion et al. (2003), Gray et al. (2013) bring our attention to three sources of strategic value in digital transformation: value chains to create new products and business processes, channels to transmit solutions to customer problems (client interaction), and networks for client-partner interaction through mediation technologies, i.e. platforms, consistent with the product portfolio perspective. All these sources involve digital interactions with external stakeholders (the edge), including clients. Based on the client value-generating interaction model that the business organization decides, impacts can range from operational processes and the organizational structure, to the transformation of the business model (Gray et al., 2013), both influencing the center and at the edge of the organization.

Without being comprehensive, typical digital transformation strategies that relate to product portfolio are two-fold, involving allegedly superior, even personalized, customer experience and the development of new products and services that are enriched or digitized (Sanchez, 2017). These strategies, according to the three categories proposed by Andal-Ancion et al. (2003), have the potential to increase resilience. In other words, resilience efforts can implicate the integration of digital technologies and use the framework shown in Figure 1 to determine actions that digitalize product portfolios and generate new value-added as part of a digital business model.

## 3. ORGANIZATIONAL RESILIENCE

The framework presented in Figure 1 aims to establish the actions that shape the development of digital transformation in a given organization. From a resilience perspective, digitalization would be expected to bring more client interaction alternatives, and lead to an update of processes and organizational structure, particularly among SMEs, which abound in sectors highly affected by COVID-19, such as retail and services (Gregurec, Tomičić Furjan, & Tomičić-Pupek, 2021). Eventually, digitalization for resilience would initiate the transformation of the business, involving product portfolio and client interactions, that is, the value proposition. However, to reach transformation, Gregurec et al. (2021) identified that organizational, financial, social and customer-driven changes are more relevant than technology. In other words, SMEs tend to use social media, platforms and mobile technology as a mean to reach customers and facilitate service delivery, which are easier to use and enable enhanced value propositions. Coincidentally, since SMEs still lack the necessary skills to cope with digital complexity, it has been observed that main responses among SMEs involve digital actions such as social media and online selling, e-banking usage, online community participation and home office work (Klein & Todesco, 2021).

Several models have tried to explain the main features of organizational resilience. For instance, Annarelli et al. (2020) identify static and dynamic characteristics that lead to organizational resilience, whose dimensions include adaptability, reliability, agility, effectiveness, flexibility and recovery. In contrast, Xiao & Cao (2017) conceptualizes resilience as a transition process, which starts at the individual level, whose aggregated characteristics are passed to the team, to eventually conform the organizational perception and behavior that define its resilience. Patriarca et al. (2018) propose a grid that combines learning, monitoring, responding and anticipating as key elements in organizational resilience. Similarly, Denyer (2017) has integrated a model that considers four organizational resilience tensions, namely, performance optimization (improving and exploiting), preventative control (monitoring and complying), mindful action (noticing and responding), and adaptive innovation (imagining and creating). All these models identify high-level attributes and actions that organizations should progressively adopt in order to gain resilience. Undoubtedly, resilience requires capabilities to predict scenarios and adapt, leading to short term responses and long term transformation (Miceli et al., 2021). However, little consideration has been granted to the identification of specific context-dependent triggers, which combine with pre-existing sociotechnical configurations (Tsoutsos & Stamboulis, 2005), to define digital transformation actions (object of impact) in order to gain such resilience. Additionally, digitalization allows for a continuous connection of all stakeholders, both internally and externally, to maintain business processes in operation (Miceli et al., 2021). Thus, digital transformation requires the participation of a variety of players (Steward, 2012), including the ecosystem of stakeholders. The three categories of DT shown in Figure 1 constitute digital transformation actions that can be explained by contextual variables, this being the research problem that this paper aims to explore.

## 4. APPROACH/METHOD

Drawing from case-oriented techniques –Qualitative Comparative Analysis (QCA) – this paper aims to derive necessary and sufficient conditions of contextual variables related to COVID-19 crisis, namely layoffs, staff income and benefits reduction, shortages in the supply chain, income decline, demand losses, cash flow scarcity or limited access to financial services, that may have led to digital transformation actions, particularly online client interaction and new product development (see Figure 2 below). Based on set theory, QCA evaluates logical Boolean combinations of conditions (equivalent to dependent variables) to determine associations to a predetermined outcome (equivalent to independent variable) (Ragin, 1987). Additionally, standard multivariate statistics will be used to supplement QCA results and provide deeper explanatory power. Datasets will be taken from Mexico's National Institute of Statistics (INEGI). Sample size is 1564 SMEs.

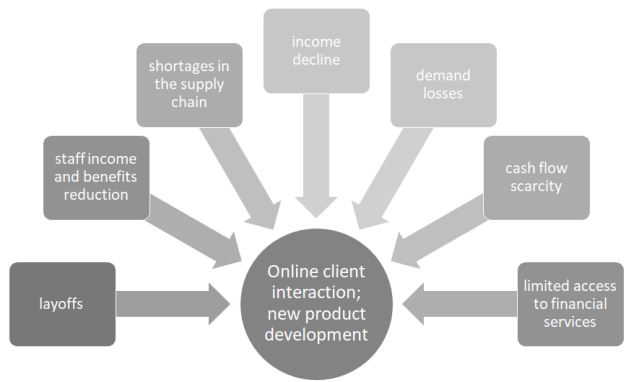

**Figure 2. Potential causal conditions and outcomes**

On the other hand, to gain knowledge about relations with the outside environment (communities, networks or ecosystems) that might lead collaborative actions in order to gain cluster resilience, information would be captured from interviews and questionnaires applied to selected industry associations in the country (to start in the Fall, 2021), which may provide a first glance of DT-resilience group implications.

## 5. VALUE/ IMPLICATIONS

This framework proposed is still a general model and its components can still be broken down in further constituents, according to the business characteristics and structure of a given organization. However, it invites scholars and the business organization to include causal thinking in the digital transformation design for organizational resilience, which may help also address some of the problems originated by COVID-19 in the Global South. As indicated by Klein & Todesco (2021), SMEs that expect to survive would need to adapt by bootstrapping immediate digital actions, learn and prepare for the expected new normal. Hence, this research aims to identify sufficient and/ or necessary conditions that should be present for at least one of the DT dimensions to occur, posing theoretical questions about crisis conditions that encourage a particular form of DT initiative. Finally, group actions may indicate the importance of clusters and networks to gain resilience in participating enterprises.

# REFERENCES AND CITATIONS